\documentclass[twocolumn]{aastex63}

\usepackage{verbatim}

\received{\today}
\revised{\today}

\submitjournal{Research Notes of the AAS}

\shorttitle{Evaluating and enhancing candidate clocking systems for CHIME/FRB VLBI Outriggers}

\shortauthors{Cary et al.}
\graphicspath{{./}{figures/}}

\begin{document}


\title{Evaluating and Enhancing Candidate Clocking Systems for CHIME/FRB VLBI Outriggers}

\correspondingauthor{Savannah Cary}
\email{scary@wellesley.edu}

\author[0000-0003-1860-1632]{Savannah Cary}
\affiliation{Department of Astronomy, Wellesley College, 106 Central Street, Wellesley, MA 02481, USA}
\affiliation{MIT Kavli Institute for Astrophysics and Space Research, Massachusetts Institute of Technology, 77 Massachusetts Ave, Cambridge, MA 02139, USA}

\author[0000-0002-0772-9326]{Juan Mena-Parra}
\affiliation{MIT Kavli Institute for Astrophysics and Space Research, Massachusetts Institute of Technology, 77 Massachusetts Ave, Cambridge, MA 02139, USA}
\author[0000-0002-4209-7408]{Calvin Leung}
\affiliation{MIT Kavli Institute for Astrophysics and Space Research, Massachusetts Institute of Technology, 77 Massachusetts Ave, Cambridge, MA 02139, USA}
\affiliation{Department of Physics, Massachusetts Institute of Technology, 77 Massachusetts Ave, Cambridge, MA 02139, USA}

\author[0000-0002-4279-6946]{Kiyoshi Masui}
\affiliation{MIT Kavli Institute for Astrophysics and Space Research, Massachusetts Institute of Technology, 77 Massachusetts Ave, Cambridge, MA 02139, USA}
\affiliation{Department of Physics, Massachusetts Institute of Technology, 77 Massachusetts Ave, Cambridge, MA 02139, USA}
\author[0000-0003-4810-7803]{J.F. Kaczmarek}
\affiliation{Dominion Radio Astrophysical Observatory, Herzberg Research Centre for Astronomy and Astrophysics, National Research Council Canada, PO Box 248, Penticton, BC V2A 6J9, Canada}

\author[0000-0003-2047-5276]{Tomas Cassanelli}
\affiliation{David A.~Dunlap Department of Astronomy \& Astrophysics, University of Toronto, 50 St.~George Street, Toronto, ON M5S 3H4, Canada}
\affiliation{Dunlap Institute for Astronomy \& Astrophysics, University of Toronto, 50 St.~George Street, Toronto, ON M5S 3H4, Canada}

\collaboration{6}{(CHIME/FRB Collaboration)}



\begin{abstract}

As the Canadian Hydrogen Intensity Mapping Experiment (CHIME) has become the leading instrument for detecting Fast Radio Bursts (FRBs), CHIME/FRB Outriggers will use very-long-baseline 
interferometry (VLBI) 
to localize FRBs with milliarcsecond precision. The CHIME site uses a passive hydrogen maser frequency standard in order to minimize localization errors due to clock delay. However, not all outrigger stations will have access to a maser. This report presents techniques used to evaluate clocks for use at outrigger sites without a maser. More importantly, the resulting algorithm provides calibration methods for clocks that do not initially meet the stability requirements for VLBI, thus allowing CHIME/FRB Outriggers to remain true to the goal of having milliarcsecond precision. 

\end{abstract}



\section{Introduction}
\label{sec:intro}
To localize Fast Radio Bursts \citep[FRBs, millisecond-long radio bursts originating from distant galaxies,][]{lorimer2007bright}, the
CHIME/FRB Outriggers project is using VLBI with telescopes situated across North America. VLBI localizes its sources by recording the signals' arrival times between two or more telescopes and measuring the propagation delay. Imperfections in a VLBI station's internal clock leads to undesired contributions to the measured delay; this report focuses on mitigating this timing offset (referred to as ``clock delay"). Outrigger stations will rely on well-localized, bright sources suitable for phase calibration on large baselines (referred to as calibrators), alongside a local frequency reference (i.e. clock) that is stable over timescales for which there are no calibrators in the field of view (FOV). 

Our goal is to correct for unwanted delay due to clock errors, ultimately creating a system where timing errors do not exceed 200 ps on timescales of at least 1000\,s \citep{Mena_2021}. We therefore have created a data analysis pipeline to determine a clock's performance against our specification for CHIME/FRB Outriggers. Our pipeline handles measured and simulated clock data. 

In this note we present our methods for evaluating candidate clocking systems for VLBI. Using measured data from experimental tests, we then analyze the performance of the EndRun Technologies Meridian II US-Rb rubidium oscillator and determine if this clock can be used at outrigger sites. 

\section{Evaluating Clock Performance}
\label{sec:pipeline}
\subsection{Providing a timestream}

Testing clock performance requires a timestream that represents the system's timing errors. This algorithm utilizes real and mock timestreams, where the mock timestreams are simulated based on Allan deviations provided by the clock manufacturer. Because the instability of most clocks can be modeled as a sum of power-law type noises \citep{Interferometry_Thompson}, it is assumed that the Allan variance follows a linear combination of the following power-law type noises: white phase modulation $(\sigma_y \propto \tau^{-1})$, white frequency modulation $(\sigma_y \propto \tau^{-1/2})$, flicker frequency modulation $(\sigma_y \propto \tau^0)$, and random walk frequency modulation noise $(\sigma_y \propto \tau^{1/2})$.

\subsection{Interpolation Techniques}
\label{sec:interpolating}

Calibrators are used to infer clock timing offsets. For our analysis, a calibrator is a reliable delay measurement. We can run the recorded or simulated measurement of clock delays through an interpolation algorithm that simulates the presence and absence of calibrators in the FOV; the timestream is masked for periods corresponding to when there is no observable calibrator. The chosen clock must be stable on the timescales spanning between calibrator observations. 

To evaluate performance based on varying separation times between calibrators, we interpolate between observations of the mock calibrators to estimate the masked clock delay and compare to true delay.
Two mock calibrators are placed at a random section of the timestream with a separation time $\tau_\mathrm{{sep}}$ between the end of the first transit and beginning of the second. This is motivated by the fact that CHIME and the Outriggers are stationary instruments, relying on Earth's rotation to survey the sky. 
Calibrators are observable for 540~s, which corresponds to the transit time through the CHIME FOV at zenith. 
Data from these 540\,s are ``recorded'' with a chosen simulated integration period of $\tau_{\mathrm{int}}$, providing 540\,s/$\tau_{\mathrm{int}}$ data points collected per calibrator. All clock delay measurements within this span of $\tau_{\mathrm{int}}$ are averaged and considered to be the recorded delays associated with the integration period. 
The timestamp associated with each sample is taken to be the middle-time of each integration. These data points, 540\,s$/\tau_{\mathrm{int}}$ for each calibrator, the basis for inferring the measurements of timing offsets of the timestream. 

Because calibrators are not infinitely bright, we expect uncertainty with each measurement of clock delay. Varying degrees of calibration errors, $\sigma_{\mathrm{cal}}$, are added to each data point via a normal distribution to make the simulation as realistic as possible. The calibration error is based off varying the calibrator's signal-to-noise ratio (SNR). Other noise contributions are the noise generated when injecting the clock signal into the telescope's correlator, $\sigma_{\mathrm{transfer}}$, and can be calculated from the timestream itself.
For simulated timestreams, this noise is added. 

\begin{figure*}
    \centering
    \includegraphics[width = \textwidth]{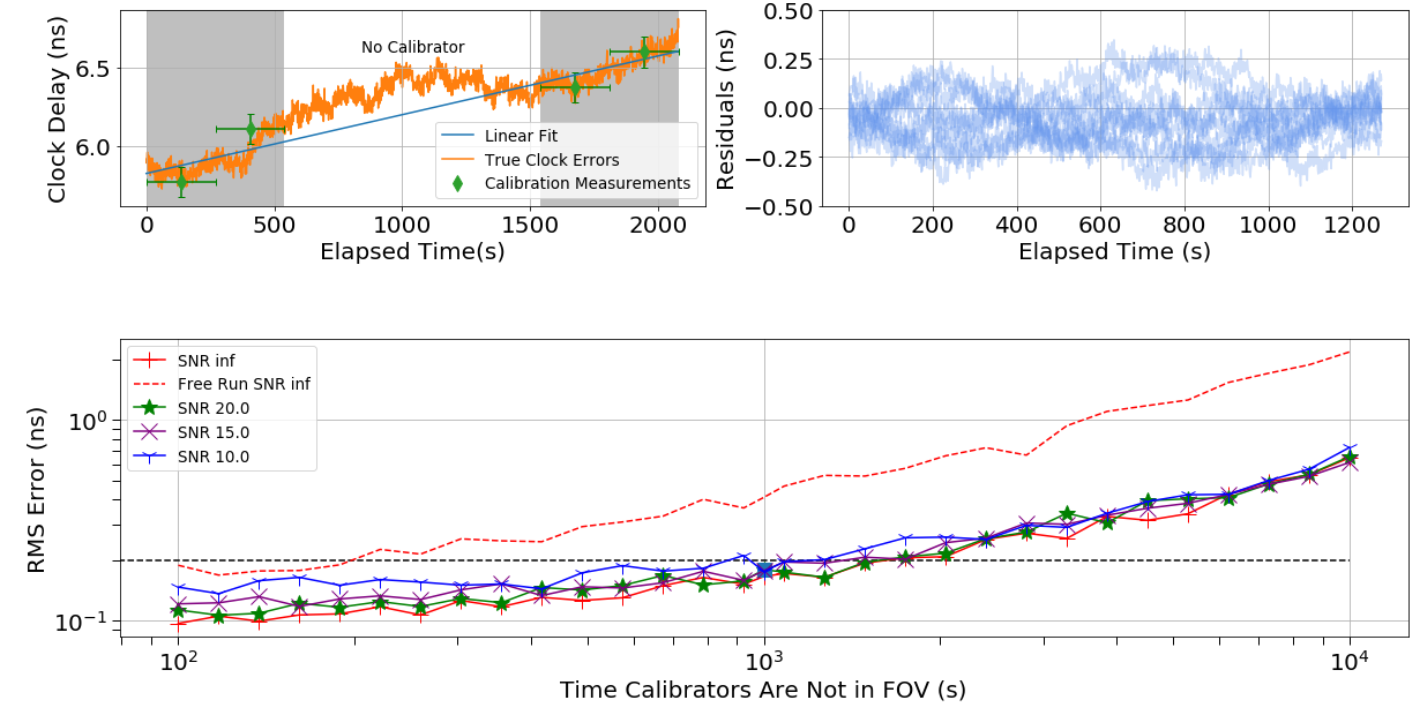}
    \caption{Simulated calibration procedure using measured timing offsets for a rubidium clock. The top left graph is a single iteration of the process described in Section~\ref{sec:interpolating} for a timestream with noise added according to an SNR of 20. Two calibrators are simulated to be in the FOV for 540\,s (shown in the shaded area) with a separation time $\tau_{\mathrm{sep}}$ 1000\,s. The green diamonds are the unmasked data, with the horizontal error bars being the integration period per calibrator. The vertical error bars represent the calibration error and clock transfer noise $(\sigma_{\mathrm{cal}}^2 + \sigma_{\mathrm{transfer}}^2)^{1/2}$. The upper right plots the residuals from interpolating 10 out of 100 iterations for a $\tau_{\mathrm{sep}}$ = 1000\,s. The corresponding RMS error for this specific instance is marked by the blue square in the bottom panel. Shown also is how well we expect the clock to perform given different values of noise given via SNRs of 10, 15, 20, and infinite. The horizontal dashed line represents the desired threshold for accounting for clock delay: 200~ps. Plotted is also the result of the calibration if no interpolation is applied, labeled ``Free Run''. By implementing this algorithm, at infinite SNR and 1000\,s, the clocking errors are reduced by over a factor of $\sim$3.}
    \label{fig:setup}
\end{figure*}

In Fig.~\ref{fig:setup}, the top left plot is an example of this masking method for simulating two calibrators. 
We interpolate between calibrations using the best fit line obtained from the collected calibration points.

For this clock, a linear fit provided the lowest RMS uncertainty. However, our methods can handle other forms of fitting functions, such as cubic, quadratic, and general cross validation smoothing splines \citep{Craven1978}, dependent on what is best for a given clock's performance.

Each delay is weighted 
based on the inverse variance $1/\sigma^2$; $\sigma$ is determined from the calibration error and transfer noise and equals $(\sigma_{\mathrm{cal}}^2 + \sigma_{\mathrm{transfer}}^2)^{1/2}$. The best fit function is what would be used to account for clock delay when phasing the array.  

The process is repeated for several random sections of the timestream. The top right plot of Fig.~\ref{fig:setup} has the resulting residuals for a single $\tau_{\mathrm{sep}}$. Simulating the worst-case scenario, the RMS is calculated using the maximum residual per iteration between the fitting function and the timestream. The lower plot of Fig.~\ref{fig:setup} represents having run multiple random
interpolations across the timestream, per desired $\tau_{\mathrm{sep}}$.

\section{results from the rubidium oscillator}
\label{sec:meridian}
 
The results in Fig.~\ref{fig:setup} are from the rubidium oscillator, a clock now proved to be more cost effective compared to a maser.
Without use of our algorithm, at infinite SNR the clock meets stability requirements for $\tau_{\mathrm{sep}}$~\textless~200\,s; however, after applying our interpolation methods, the clock meets our requirements at times $\geq$ 1000\,s for SNR \textgreater 10. 

In conclusion, these methods make it possible to evaluate clocks for stations without a maser for CHIME/FRB Outriggers. These methods also make it possible to extend the stability of sub-optimal high-precision timing standards, allowing for a more cost-effective solution to accounting for clock errors. In the future, this analysis can be implemented where frequency standards are not originally stable enough on their own, enabling the CHIME/FRB Outriggers VLBI network to meet its strict timing and localization requirements.



\vspace{5mm}

\software{\texttt{AllanTools}~\citep{2018ascl.soft04021W}}




\bibliography{references}{}

\bibliographystyle{aasjournal}

\end{document}